\documentclass[10pt,conference]{IEEEtran}
\usepackage{cite}
\usepackage{amsmath,amssymb,amsfonts}
\usepackage{algorithmic}
\usepackage{graphicx}
\usepackage{textcomp}
\usepackage[hyphens]{url}
\usepackage{fancyhdr}
\usepackage{hyperref}
\usepackage[dvipsnames]{xcolor}
\usepackage{soul}
\usepackage{caption}
\DeclareCaptionLabelFormat{cont}{#1~#2\alph{ContinuedFloat}}
\usepackage{verbatim}

\usepackage{graphicx,pifont} 
\let\oldding\ding 
\renewcommand{\ding}[2][1]{\scalebox{#1}{\oldding{#2}}} 
\usepackage{listings}

\title{Targeted Wearout Attacks in Microprocessor Cores}

\def\hpcacameraready{} 
\newcommand\hpcaauthors{Joshua Mashburn$\dagger$, Johann Knechtel$\ddagger$, Florian Klemme$\ast$, Hussam Amrouch$\star$, Ozgur Sinanoglu$\ddagger$, and Paul Gratz$\dagger$}
\newcommand\hpcaaffiliation{Texas A\&M University$\dagger$, New York University Abu Dhabi$\ddagger$, Independent$\ast$, TU Munich$\star$}
\newcommand\hpcaemail{joshualmashburn@tamu.edu, johann@nyu.edu, klemme@iti.uni-stuttgart.de, amrouch@tum.de, ozgursin@nyu.edu, pgratz@tamu.edu}

\author{
  \ifdefined\hpcacameraready
    \IEEEauthorblockN{\hpcaauthors{}}
      \IEEEauthorblockA{
        \hpcaaffiliation{} \\
        \hpcaemail{}
      }
  \else
    \IEEEauthorblockN{\normalsize{HPCA \hpcayear{} Submission
      \textbf{\#\hpcasubmissionnumber{}}} \\
      \IEEEauthorblockA{
        Confidential Draft \\
        Do NOT Distribute!!
      }
    }
  \fi 
}

\begin{document}
\maketitle

\ifdefined\hpcacameraready 
  \thispagestyle{plain}
  \pagestyle{empty}
\else
  \thispagestyle{plain}
  \pagestyle{plain}
\fi

\newcommand{\hpcaheight}{0mm}
\ifdefined\eaopen
\renewcommand{\hpcaheight}{12mm}
\fi


\begin{abstract}
Negative-Bias Temperature Instability is a dominant aging mechanism in nanoscale CMOS circuits such as microprocessors.  With this aging mechanism, the rate of device aging is dependent not only on overall operating conditions, such as heat, but also on user controllable inputs to the transistors.  This dependence on input implies a possible timing fault-injection attack wherein a targeted path of logic is intentionally degraded through the purposeful, software-driven actions of an attacker, rendering a targeted bit effectively stuck.  

In this work, we describe such an attack mechanism, which we dub a "\textbf{Targeted Wearout Attack}", wherein an attacker with sufficient knowledge of the processor core, executing a carefully crafted software program with only user privilege, is able to degrade a functional unit within the processor with the aim of eliciting a particular desired incorrect calculation in a victim application.  Here we give a general methodology for the attack. We then demonstrate a case study where a targeted path within the fused multiply-add pipeline in a RISC-V CPU sees a $>7x$ increase in wear over time than would be experienced under typical workloads.  We show that an attacker could leverage such an attack, leading to targeted and silent data corruption in a co-running victim application using the same unit.
\end{abstract}

\section{Introduction}



Very Large Scale Integration (VLSI) faces complex design challenges in delivering high performance while managing power consumption. Historically, Moore's Law and Dennard scaling addressed these demands through feature size
reduction. However, ever-shrinking process nodes introduce significant reliability hurdles for Complementary Metal-Oxide-Semiconductor (CMOS) logic. Among others, Song et al.~\cite{songAmdahlsLawLifetime2016} 
note that reliability is now an increasingly critical design constraint alongside power, performance, and area.



A key concern for reliability is device aging, where components degrade over time, eventually limiting the operational lifetime of a circuit. Several physical mechanisms contribute to this degradation in
Metal-Oxide-Semiconductor Field-Effect Transistors (MOSFETs), the building blocks of CMOS logic. Among the most prominent are Negative-Bias Temperature Instability (NBTI) and Hot-Carrier Injection (HCI)~\cite{alamComprehensiveModelPMOS2007}. Both phenomena
can increase the threshold voltage ($V_{th}$) of transistors, consequently increasing their switching delay. As these delays accumulate, critical paths within the circuit eventually fail to meet timing requirements,
leading to incorrect operations and effective failures.


The fact that reliable operation of VLSI circuits is subject to aging also holds ramifications for hardware security.
First, note that Fault Injection Attacks (FIAs) are a well-established threat that
allows adversaries to corrupt data or disrupt operation~\cite{breierHowPracticalAre2022,karimiMAGICMaliciousAging2015,kimFaultsInjectionMethods2007,tangCLKSCREWExposingPerils2017a}. 
On the one hand, typical FIAs employ hands-on techniques like clock glitching, laser fault injection, etc., which require physical access, specialized equipment, and thorough knowledge of the device under attack.
On the other hand, excessive aging can enable timing-based FIAs without direct physical access.
Aging mechanisms like NBTI directly depend on user controllable activity; thus, aging can be deliberately accelerated through software efforts.
This is the key premise of our work.
Finally, while traditional mitigation techniques such as timing guardbands~\cite{kimAgingGracefullyApproximation2019a} can limit the impact of aging, they are focused on regular wearout, not on maliciously accelerated
wearout of particular paths. Thus, modern VLSI circuits are left vulnerable to such attacks.


\begin{figure*}
    \centering
    \includegraphics[width=1.0\linewidth]{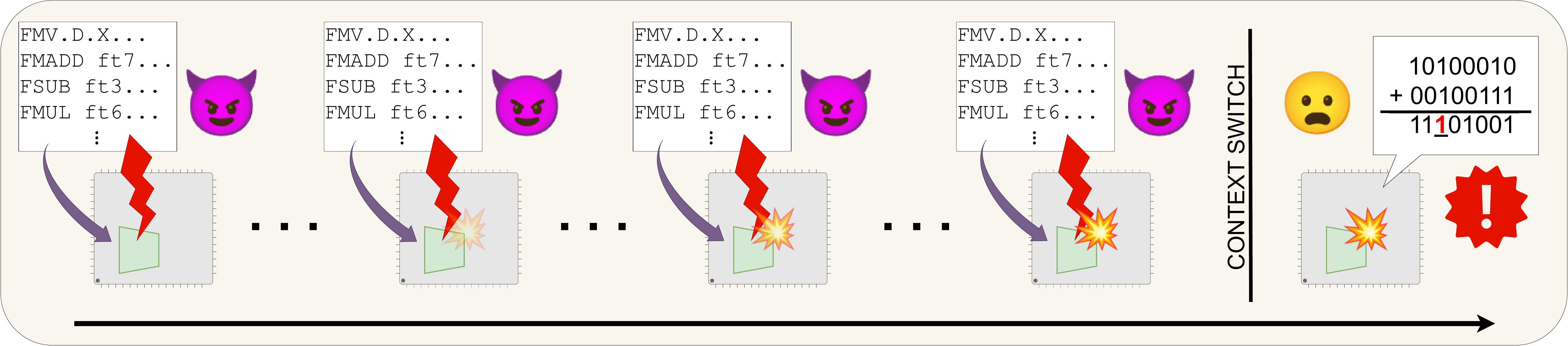}
    \caption{Working principle of the proposed wearout attack. An adversarial user runs carefully crafted user-level code aimed to excessively accelerate aging on selected logic paths in a circuit under attack.
	    This maliciously triggered wearout subsequently induces incorrect computation results / data corruption for any benign user of the same circuit.}
    \label{fig:introcartoon}
\end{figure*}

In this work, we show that an attacker, by crafting specific input patterns, can indeed induce \textit{targeted} aging on pre-chosen logic paths within VLSI circuits, even in the presence of guardbands, effectively inducing
timing faults remotely.
This represents a stealthy attack vector capable of causing malfunctions, data corruption, or denial-of-service without requiring physical access.
We introduce \textbf{Targeted Wearout Attacks (TWA)}, a class of attacks that leverage these principles.
A high-level overview is illustrated in Figure \ref{fig:introcartoon}.

More specifically, we propose a methodology to identify vulnerable paths within microprocessor cores and systematically generate input patterns designed to accelerate NBTI-induced aging along those specific paths.
Our approach analyzes gate-level netlists and employs Automatic Test Pattern Generation (ATPG) and formal methods to carefully devise these malicious inputs.
\sethlcolor{green}
In 2015, Karimi et al.~\cite{karimiMAGICMaliciousAging2015} demonstrated the feasibility of malicious aging in a processor core. That work targeted the most critical path of an in-order core, causing general failure.  The TWA approach provides more flexibility by targeting non-critical-path bits, and by focusing on data paths on a core, it produces a silent, directed data corruption attack.  In contrast to the prior work, the processor ostensibly continues functioning as normal while the victem's data is corrupted in a controlled way according to the designs of the attacker.  Further, because all NBTI wear is somewhat reversible, it is possible that the attacker could modulate the attack to only be in effect for a limited, specific time.
\sethlcolor{yellow}
To demonstrate the feasibility and impact of TWA, we present a range of experiments targeting the floating-point unit (FPU) of a popular open-source RISC-V microprocessor core. Using our
proposed methods, we show that a carefully constructed TWA can accelerate the aging of the target path, reducing its functional lifetime by up to a factor of 7.1 compared to nominal operation under standard benchmark
workloads.

Major contributions of this work include:
\begin{itemize}
\item The conceptualization and demonstration of TWA, a stealthy, software-based attack leveraging transistor aging to induce timing faults and resulting threats like data corruption.
\item A framework for analyzing VLSI designs to identify susceptible paths and generate effective TWA input patterns using both ATPG and formal methods.
\item Thorough assessment of the attack's effectiveness by a range of experiments on a modern RISC-V FPU, demonstrating significant lifetime reduction of a targeted path.
For practical relevance, we utilize aging-characterized technology libraries and standard guardbands.
\end{itemize}

This paper is organized as follows.
Section~\ref{sec:BG} provides background on CMOS aging and mitigation techniques, on FIAs, and on ATPG and formal methods.
Section~\ref{sec:TWA} introduces the TWA approach, with its threat model and the general attack flow.
Section~\ref{sec:CS} describes our RISC-V FPU case study in detail, covering both the technical setup and the general methodology.
Sections~\ref{sec:results} and~\ref{sec:analysis} present and analyze the results.
Section~\ref{sec:disc} discusses potential mitigation strategies and future work,
	and Section~\ref{sec:concl} concludes the work.

\section{Background}
\label{sec:BG}



\subsection{Device and Circuit Aging}

As CMOS circuits age, the operating parameters of individual transistors shift until the circuit can no longer reliably perform correctly within acceptable tolerances.
There are two primary aging effects for MOSFETs: HCI and BTI. Both increase $V_{th}$, reduce drain currents, and thus increase the switching delays, though they differ in their causes and 
impacts on PMOS and NMOS transistors.

HCI is a phenomenon in which charge carriers pass through the channel of a MOSFET with high kinetic energy, so some of them tunnel into the gate oxide, becoming trapped. The related interference reduces carrier mobility
and the transconductance. This mechanism is more prevalent in PMOS transistors and is highly dependent on a high switching activity factor to induce a strong electric field.
For BTI, charge carriers become trapped along the Si-SO\textsubscript{2} interface, increasing $V_{th}$ and reducing the drain current~\cite{alamComprehensiveModelPMOS2007}.
For PMOS transistors, the stuck carriers are holes, and the phenomenon is referred to as negative-bias temperature instability or \textit{NBTI}.
The analogue for NMOS transistors is positive-bias temperature instability or \textit{PBTI}.

NBTI is considered the most pervasive reliability issue in CMOS logic, persisting from the early SiO\textsubscript{2} devices, to silicon oxynitride (SiON) dielectric devices, to today's high-k Metal Gate (HKMG) planar
MOSFETs and FinFETs~\cite{mahapatraReviewNBTIMechanisms2018}. Given its importance, this work focuses on NBTI.

\subsection{Logical Impact of Circuit Aging}
\label{sec:logicalimpact}

\begin{figure}
    \ContinuedFloat*
    \includegraphics[width=1.0\linewidth]{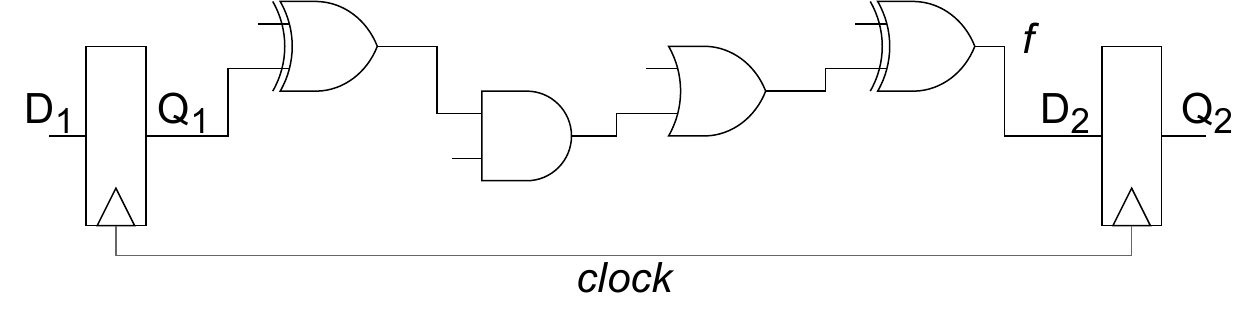}
    \caption{An exemplary logic path between two FFs.}
\end{figure}
\begin{figure}
    \ContinuedFloat
    \includegraphics[width=1.0\linewidth]{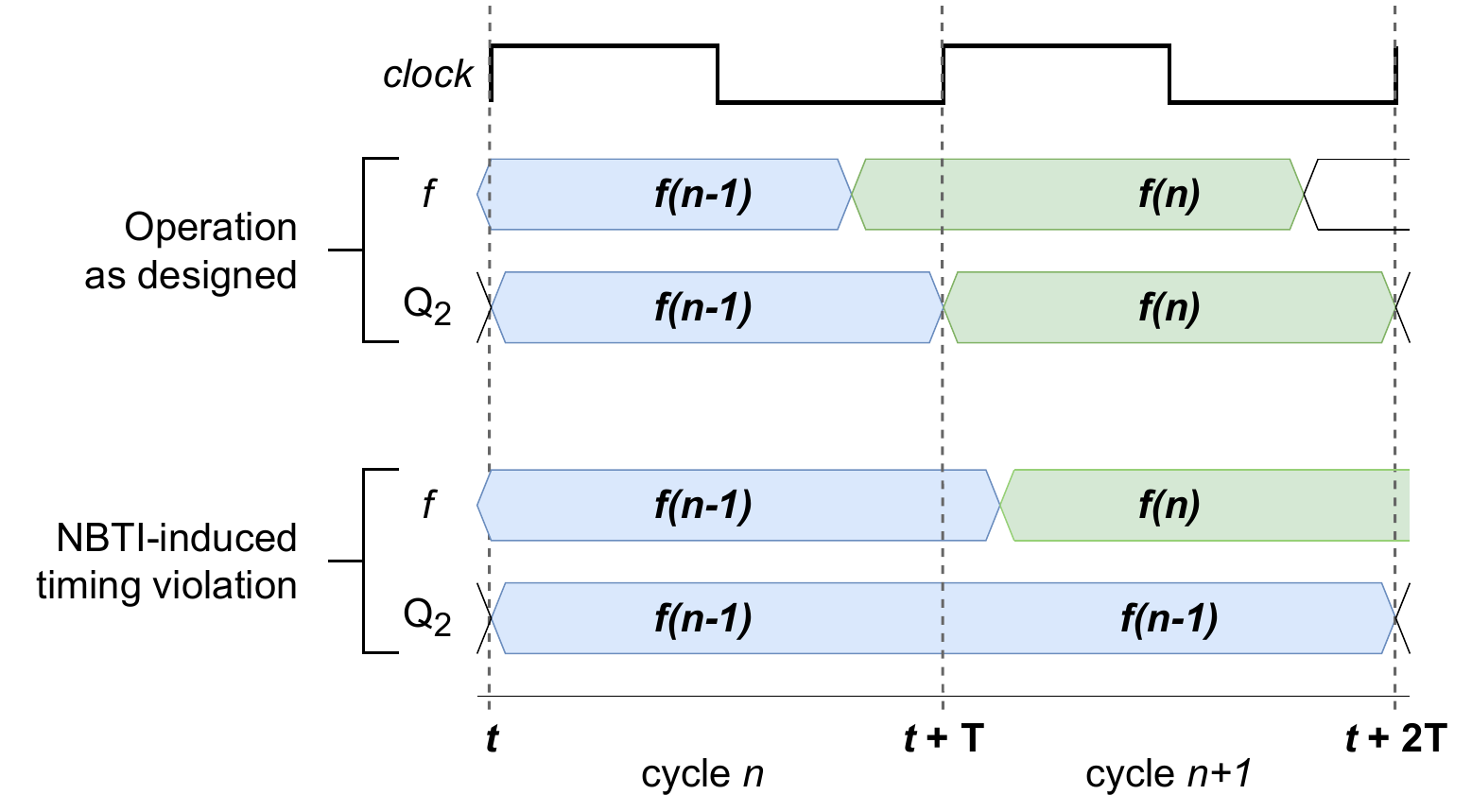}
    \caption{Exemplary waveforms illustrating the operation of this logic path without timing violations (top) vs NBTI-induced aging causing a timing violation (bottom). For the latter case, the result of $f$ does not propagate in time, so the old data $Q_2$ is falsely re-latched.}
    \label{fig:timingviolation}
\end{figure}

For a CMOS circuit to operate correctly, its input clock must operate with a long enough period $T$ to accommodate the cumulative switching delays of all the gates along the critical
path~\cite{kothawadeAnalysisIntermittentTiming2012}.

Figure~\ref{fig:timingviolation}a provides a simple example for logic paths: this path connects four gates between two flip-flops (FFs) and operates at a frequency of $\frac{1}{T}$.
For the second FF to latch the correct output of that path, here called $f$, $T$ must be greater than the sum of the delays incurred by each gate along the path, plus the setup time
for the FF itself~\cite{nunesBTIHCITDDB2013}.
Figure~\ref{fig:timingviolation}b compares two timing scenarios. In the first case, during cycle $n$, the output signal of the path finishes propagating to the D pin of the second FF in time for the rising clock edge, so that in
cycle $n+1$, the output pin Q2 now holds the correct signal. In the second case, however, NBTI aging has expanded the delays of the gates such that the result of $f$ in cycle $n$ does not propagate to the second FF in
time for the rising clock edge, causing a timing violation. That is, in cycle $n+1$, the second FF fails to latch the new value of $f$ and thus continues to propagate the previous value, and the path is considered to be
in a failure state.
Naturally, any such failures will lead to data corruption, disruption of operation flows, hiccups in state management, etc.

\subsection{Modeling of NBTI Aging}


Predicting long-term NBTI degradation in a transistor follows the Reaction-Diffusion (R-D) model, more specifically to derive the shift in $V_{th}$ due to NBTI stress and relaxation periods. This shift in $V_{th}$ degrades the transistor delay following the Alpha-Power Law~\cite{sakuraiAlphapowerLawMOSFET1990}:
\begin{equation} \label{eq:alphapowerlaw}
    d_g \propto \frac{V_{dd}}{\mu(V_{dd}-V_{th})^\alpha}
\end{equation}
where $d_g$ is the gate transition delay. Note that $\mu \propto T^-1.5$ where T is temperature, and $\alpha$ = 1.3.

We use the models derived by Lu et al.~\cite{luStatisticalReliabilityAnalysis2009} and Kim et al.~\cite{kimUseItLose2015} to calculate the delay adjustment for a path of $N$ gates at time $t$:
\begin{equation} \label{eq:gatedelay}
\resizebox{1\linewidth}{!} 
{
    \ensuremath{\Delta d(t) = \sum_{i=0}^{N-1} \Delta d_i(\beta_i,t) = A \times e^{-\frac{nE_{aNBTI}}{kT}} \times t^n \times \sum_{i=0}^{N-1} \left(\frac{\beta_i}{1-\beta_i}\right)^n}
}
\end{equation}
where $\beta_i$ is the duty cycle of gate $i$, $A$ is a fitting constant, $E_{aNBTI}$ is the activation energy, $k$ is Boltzmann's constant, $T$ is the temperature, and $n$ is the time exponent, approximately 1/6~\cite{alamComprehensiveModelPMOS2007,mahapatraReviewNBTIMechanisms2018}.

Kim et al.~\cite{kimUseItLose2015} introduced the concept of an aging-related \textit{Acceleration Factor}, i.e., a ratio of path lifetimes derived from Equation \ref{eq:gatedelay},
to compare the critical paths of two systems.
Importantly, this factor is not to be confused with acceleration of operation etc.
The factor is defined as follows:
\begin{equation} \label{eq:accelerationfactor}
    AF(x)=\frac{T_{lifetime}(x)}{T_{lifetime}(ref)}=\left( \frac{\sum\limits_{j=0}^{M-1}\left(\frac{\beta_j}{1-\beta_j}\right)^n} {\sum\limits_{i=0}^{N-1}\left(\frac{\beta_i}{1-\beta_i}\right)^n} \right)^{1/n}
\end{equation}
where $T_{lifetime}(x)$ is the useful lifetime of a Device Under Test (DUT), be it for some NBTI mitigation or some workload of interest, while $T_{lifetime}(ref)$ is the lifetime of a similar device under a baseline workload.
Therefore, $\beta_i$ is the duty cycle of gate $i$ along the critical path in the system under consideration, while $\beta_j$ is the duty cycle of gate $j$ along the critical path in the reference system.

\subsection{Mitigation of Aging}


Various techniques to monitor and mitigate $V_{th}$ degradation caused by NBTI stress have been proposed, all with varying costs, power, and area overheads.
Importantly, none of the established techniques described below consider maliciously targeted and accelerated aging, leaving circuits vulnerable to our proposed TWA threat.\footnote{%
For example, guardbands can be overcome by excessive aging acceleration, as we also show
in our experiments later on. Monitors cover only selected paths, typically the most critical ones, whereas our approach can attack various paths  of choice.}

A traditional approach is to apply a reliability guardband~\cite{agarwalCircuitFailurePrediction2007,kimAgingGracefullyApproximation2019a}, i.e., increasing the slack of the critical path of a design.
This approach is simple and rather inexpensive, only impacting clock frequency. This adjustment increases the operating lifetime of the device, since the degradation must be more severe to
surpass the extended clock period.  However, this approach tends to be overly conservative, as determining the actual guardband requires activity and duty cycle data from normal and worst-case operation of the design.
In complex circuits like
CPUs, it is difficult to establish a guardband without either being too protective while sacrificing performance or being too permissive while limiting the lifetime of the circuit.
Still, notable efforts exist to improve the high-level models for aging~\cite{srinivasanCaseLifetimeReliabilityAware2004,swaminathanBRAVOBalancedReliabilityAware2017}, by calculating the average Failures-in-Time (FIT)
rate of the processor core, to better decide on the guardband.
Besides, machine learning (ML)-based techniques have been proposed to help devising an appropriately sized guardband without unnecessary performance loss. These methods characterize standard cell libraries for specified
aging conditions, such as temperature or $V_{dd}$~\cite{klemmeMachineLearningOntheFly2021,kimOnChipAgingSensor2010} or workload-dependent duty cycles~\cite{klemmeMachineLearningCircuit2021}. A designer using these libraries can more accurately
simulate the behavior of their circuit at the end of its designated lifetime under the formulated operating conditions, enabling them to more carefully select a relevant guardband.

Another inexpensive approach is to increase the operating voltage of the circuit~\cite{keaneOdomoeterCPUs2011}. This considers the Alpha-Power Law (Equation~\ref{eq:alphapowerlaw}) between the transition delay and the difference in $V_{dd}$ and $V_{th}$ in PMOS transistors. By increasing $V_{dd}$, $V_{th}$ must experience higher degradation to experience timing failure, theoretically extending the lifetime of the device. However, the adjusted $V_{dd}$ is calculated based on long-term aging, usually about ten years, and so for the majority of the typical lifespan of the device, the resulting additional power is wasted~\cite{kimAgingGracefullyApproximation2019a}. Furthermore, this increased power consumption can raise operating temperatures, thereby exacerbating most aging mechanisms including NBTI, HCI and Time-Dependent Dielectric Breakdown (TDDB).

Finally, approaches that place less importance on the actual guardband have been proposed as well. These techniques modify the behavior of the circuit under aging, e.g., by dynamically changing computational
paths~\cite{kimAgingGracefullyApproximation2019a} to achieve approximate results or by adjusting the operating parameters of the circuit.
Adaptive body
biasing~\cite{qiNBTIResilientCircuits2008} applies the latter, requiring an on-chip monitor connected to a transistor of concern. This monitor tracks and compensates the drain current of the PMOS device by adjusting its
body bias. Timing margin monitors~\cite{saiMultiPathAgingSensor2018,synopsysincSLMPathMargin} signal when a particular path comes within some threshold of the clock, allowing the system to compensate however the architect
chooses. Still, such monitors incur power and area cost, and their deployment is usually restricted to the most critical paths.
Dynamic solutions have also been proposed at the microarchitectural level. In 2007, Abella et. al introduced Penelope, an NBTI-aware CPU that inserts exercise vectors into combinational logic during idle cycles and introduces wear mitigation on memory structures~\cite{abellaPenelopeNBTIAwareProcessor2007}. However, in the context of functional unit aging, Penelope waits for idle cycles and is thus mainly anticipates normal operating wear on the device. Colt~\cite{gunadiCombatingAgingColt2010} introduced a means of wear leveling on microprocessor structures but requires inverted logic duplication for arithmetic units.
Facelift~\cite{tiwariFaceliftHidingSlowing2008} and Clotho~\cite{vitkovskiyClothoProactiveWearout2015} direct traffic across multicore processors to mitigate and mask aging effects, and Srinivasan and Bose~\cite{srinivasanExploitingStructuralDuplication2005} proposed duplication of microarchitectural structures for redundancy, extending the processor's lifetime with spare structures.

As indicated before, none of these solutions specifically address excessive aging induced by malicious activities. Still, extending these solutions toward that end may be promising for future work; we outline more
specific thoughts in Section~\ref{sec:disc}.

\subsection{Fault Injection Attacks (FIAs)}

Device reliability has emerged as a growing problem in hardware security. In 2018, Kraak et. al. demonstrated that operational disturbances in circuits such as SRAM enable denial-of-service attacks on common security primitives and highlighted the need for more work in this area.~\cite{kraakDeviceAgingReliability2018}.
In 2021, Lienig et. al. discussed circuit reliability concerns and their impacts on device security and suggested means of improvement to the electronic design automation community~\cite{lienigSecurityClosureFace2021}. It follows that intentional aging poses a real risk to the security as well as the operational integrity of a computing device.

FIAs are a well-known hardware security concern, with consequences ranging from meddlesome to severe. Traditionally, an attacker with physical access to a device leverages various glitching methods to disrupt the normal
operation of a circuit~\cite{breierHowPracticalAre2022,kimFaultsInjectionMethods2007}. For example, clock glitching inserts errant clock pulses into the device's clock tree during the execution of targeted operations with the aim of disrupting the execution state and control flow~\cite{ningModelingEfficiencyAnalysis2018}. Voltage glitching achieves similar means by shifting the supply voltage outside of normal operating bounds. Electromagnetic~(EM) fault injection is also
possible, where an attacker uses an EM pulse to induce additional currents within wires that then glitch the transistors. Likewise, laser fault injection induces additional photon-induced currents~\cite{rodriguezLLFILateralLaser2019}.

While powerful in their ramifications for secure and trustworthy computation, the above means for traditional FIAs all
require physical access, and laser FIAs even require de-lidded access to the circuit~\cite{breierHowPracticalAre2022}.
Furthermore, given the inherent randomness of the underlying physical processes, resolution/focus limitations of the utilized equipment, etc., the impact of such traditional FIAs is difficult to predict and control. It is
well-known in the field of hardware security that FIAs require significant hands-on efforts for trial and error, literally.



To address these practical challenges of traditional FIAs, so-called \textit{timing FIAs} have been demonstrated recently, which seek to manipulate the timing of a circuit to induce faulty behavior in a more
controlled/targeted manner and without physical access~\cite{tangCLKSCREWExposingPerils2017a}.
For example, an attacker can misuse the energy-efficiency scaling features in modern CPUs to simultaneously reduce the operating voltage while increasing the clock rate~\cite{tangCLKSCREWExposingPerils2017a},
or intentionally accelerate aging through crafted inputs~\cite{karimiMAGICMaliciousAging2015}.

Such attacks cause signal propagation to fail reaching their intended FF by the end of the clock period, causing stale data to latch instead; also recall Figure~\ref{fig:timingviolation}b.
Indeed, the UnTrustZone work~\cite{mahmodUnTrustZoneSystematicAccelerated2024} demonstrates this concept by leveraging an aging differential in SRAM circuits to "burn-in" prior data, leaking secrets in such a way as to bypass Arm's TrustZone enclave.
Magic by Karimi et al.~\cite{karimiMAGICMaliciousAging2015} proposed also leveraging wearout in an attack on a small in-order processor.  In that work, their attack simply makes the processor non-functional after a period of time attacking the most critical path in the entire core.  In contrast, our proposed TWA approach is (i)~stealthy -- its aim is not to break the core but instead to purposely flip some bits in a victim process's data output, (ii)~can target any	(near-critical) path of choice, and (iii)~potentially reversible since NBTI wear can be reversed somewhat, making it possible to "cover the tracks" of the attacker.
Further, note that as our attack is not limited to control logic but can attack any logic, e.g., to induce data corruption on computational paths.

\subsection{Automatic Test Pattern Generation}

ATPG is essential for post-tapeout testing of VLSI circuits~\cite{synopsysincTestMAXATPGAdvanced}.
In this process, the ATPG tool analyzes a circuit design and creates a minimal set of test patterns designed to
activate any faults that may be caused by manufacturing defects. Each pattern consists of a sequence of input vectors and each pattern covers one or more potential faults (or rather their location in the circuit).
These patterns are then applied to the inputs of the DUT,
with the intention of both activating a fault, if present, and propagating the faulty result to observable outputs.
Depending on the complexity of the circuit, thousands or more patterns have to be generated to cover as
many potential faults as possible.

\subsection{Formal Verification}

Formal verification methods are key to pre-tapeout verification of VLSI circuits.
In contrast to functional verification (testing), these methods mathematically verify correctness of a design and include model checking, equivalence checking, theorem proving, and Boolean Satisfiability solving.
Many tools exist that leverage one or more of these methods, e.g., Cadence JasperGold~\cite{cadencedesignsystemsincJasperEngineSelection2024} or Synopsys VC Formal~\cite{synopsysincVCFormalFormal}.
The use of these tools is enabled by SystemVerilog Assertions, the formal-language extension to SystemVerilog and in turn to Verilog.
For example, one can express a \textit{cover} to describe how a state or sequence of states is to be reached
in a sequential design. The tools then formally analyze the design and search over the input space for an input pattern or sequence of patterns that place the circuit in that desired state. Otherwise, they inform
that the specific state is not reachable, which is helpful for designers to debug corner cases in their logic.

\section{Targeted Wearout Attacks (TWA)}
\label{sec:TWA}

\subsection{Motivation}
\label{sec:twa_approach}

Over the lifetime of any circuit, transistors age, resulting in a degradation of their switching speeds, which can ultimately lead to the failure of the circuit by exceeding its timing guardband. 
Recall that, among other factors, NBTI-induced aging of PMOS transistors depends on the input duty cycles.
Thus, we argue that it is possible for an adversary to (i)~choose some vulnerable logic path(s) and (ii)~carefully devise and feed inputs to the path(s) so as to accelerate the rate of aging and thereby cause an
aging-induced failure of that circuit relatively early on, creating a form of timing FIA.
We call this a Targeted Wearout Attack, or TWA.

For this type of timing FIAs, we believe the computational units of a CPU core, like the FPU, would be a promising target:
\begin{itemize}
\item\textbf{Vulnerable to timing violations.} Computational units typically have long, contiguous combinatorial paths which usually are critical or near-critical.
\item\textbf{Stealthy data corruption.} Faults in these units are not as likely to lead to an immediate failure of the CPU's general functionality, thus allowing the attacker to silently corrupt the data of the victim.  This is as opposed to attacking, for example, the instruction selection logic which would quickly lead to complete system failure.
\item\textbf{Unrestricted access.} Attackers can manipulate the inputs of targeted paths in these units by simply executing their carefully devised instructions, which leverage those functional units without need for any elevated privilege.
\item\textbf{Shared resources.} Since these units are shared across the system, such attacks could even be executed from within a virtual machine, and extend attacks to orthogonal virtual machines or even the kernel or hypervisor. 
\end{itemize}

To further motivate TWAs, imagine the following attack scenarios, which are by no means exhaustive:
\begin{itemize}
    \item\textbf{Co-running VMs in an on-demand cloud service.}
    Two virtual machines, one representing the attacker and the other the victim, are scheduled to share a given node in a cloud service such as AWS, while using the container orchestration system's node affinity constraints to bind the attacker to the target node~\cite{amazonwebservicesinc.PlaceKubernetesPods}~\cite{AssigningPodsNodes}. Over time, the
    attacker induces aging in the FPU unit such that the victim's calculations are silently corrupted (to the attacker's specifications).
    \item\textbf{Malicious websites.} In a client system running a browser with several open tabs, the user visits a malicious webpage that starts a JavaScript program (or others) which aims to wearout the FPU.
    This leads to subsequent corruption of any data, e.g., for a spreadsheet calculation running on the same host system.
    \item\textbf{ML inference on edge devices.} In an autonomous vehicle, the manufacturer of the vehicle's control systems uses a neural processing unit (NPU) to ensure safe operation. To bypass these safeguards, a malicious user sideloads a program which ages the NPU, disrupting inference operations.
\end{itemize}
Again, these are possible ``use cases'' for attackers, meant to emphasize the wide-scale impact and notable severity of TWAs. In the remainder of this work, including our experiments, we focus on the actual attack implementation and evaluation in general,
	which goes without the need to focus on any particular use case.


\subsection{Threat Model}


We assume the following capabilities for the adversary:
\begin{enumerate}
    \item is skilled with VLSI tools, in particular ATPG, formal verification, and gate-level timing simulation;
    \item has access to a gate-level netlist and timing information of the circuit under attack, either obtained from the design house or by reverse-engineering the circuit and its documentation; As source-available hardware such as open-source RISC-V cores gain traction, this requirement becomes easier, even trivial, to fulfill~\cite{masoodSecurityChallengesFaced2023}.
    \item has remote access to the system running on the circuit under attack.
\end{enumerate}
All assumptions are representative of a real-world attack setting.
Importantly, not requiring physical access is a prominent feature/benefit of TWAs (and other timing FIAs) over traditional FIAs. Furthermore, the TWA payload consists entirely of benign-appearing arithmetic operations, allowing the attack to run long-term without raising suspicion. Additionally, malware persistence and obfuscation techniques~\cite{ranaMalwarePersistenceObfuscation2021} could be deployed to further ensure inconspicuous operation while the attack degrades the hardware over time.

In line with prior art on timing FIAs~\cite{tangCLKSCREWExposingPerils2017a,karimiMAGICMaliciousAging2015}, we also assume the following scope and practical limitations for the attack:
\begin{enumerate}
    \item aims to induce faults on paths that are close to, but not limited to, being timing critical;
    \item becomes effective only during computation, i.e., when the output of the faulted path is supposed to change -- it cannot independently induce specific fault values;
    \item is enabled through maliciously crafted software running on the target system, be it compiled code or assembly code;
    \item does not require elevated privileges, only user-level access to the unit holding the targeted path.
\end{enumerate}

Finally, the attacker's objective is to accelerate NBTI-induced aging on selected paths in a computational unit, so as to induce timing faults at the output of those paths and thereby induce incorrect calculation results for the
victim.


\subsection{Attack Methodology}
\label{sec:TWAgeneralflow}


\begin{figure}
    \centering
    \includegraphics[width=0.7\linewidth]{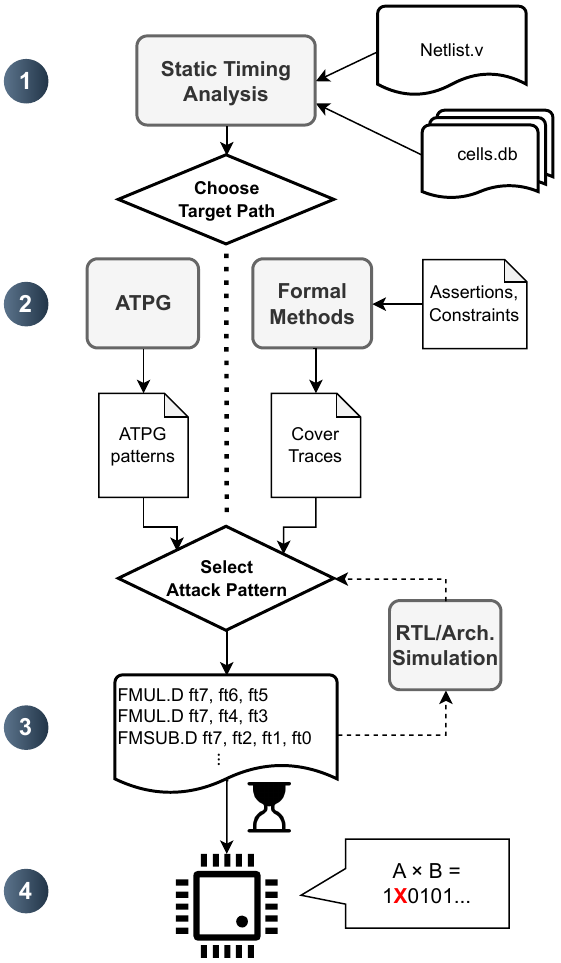}
    \caption{Process flow for TWA.
    }
    \label{fig:twaflow}
\end{figure}

Next, we describe the process and general methodology for the proposed TWA, following
the flow chart in Figure~\ref{fig:twaflow}.

\subsubsection{Timing Analysis and Path Selection}

In \ding{202}, the gate-level netlist of the unit under attack is
passed through Static Timing Analyis (STA).\footnote{%
STA is vectorless, i.e., it evaluates timing for all logic paths with their gates and interconnects and the constituting delays in general, without focus on specific input patterns, but with worst-case
assumptions across all possible signal transitions.
The result of this are delay values for all logic paths, and the path with the longest (vs shortest) delay defines the critical path in terms of the setup (vs hold) condition for FFs. In the remainder, we focus on
the setup condition only, as aging-induced delay increases can only undermine this condition but not the hold condition.}
Then, a set of paths susceptible to TWA is identified, i.e., paths within some practically tight bounds of critical timing.
Note that the specific bounds/margins are up to the attacker: while smaller bounds/margins require shorter execution of the aging-inducing malicious code until the TWA becomes effective, attackers may still consider larger
bounds/margins in case some paths of interest for FIA are less timing critical.
Once some path is chosen as target, the premise is that, upon sufficiently triggered aging and the resulting wearout,
the transition at the output of this path will have become too slow to be captured in time by connected FFs, causing an erroneous data state
(Figure~\ref{fig:timingviolation}b).

\subsubsection{Generation of Aging-Inducing Attack Patterns}
\label{sec:patterngeneration}

In \ding{203}, we derive a sequence of bit vectors that, when provided to the unit, induce aggressive NBTI aging in the PMOS transistors along the target path.
As shown in the flow chart, two complementary useful options exist to achieve this, ATPG and formal methods, as described next.

\textit{1) ATPG.}
Here we use an ATPG tool to find patterns that accelerate NBTI-induced aging by setting it to test for stuck-at-1 faults at the input legs of each gate of concern.
Testing for stuck-at-1 faults primes the circuit to (i)~drive the concerned pins to logical 0 and (ii)~propagate the concerned gates' output to test points such that it can be checked against stuck-at-1 failures while, most importantly, holding the concerned pins at logical 0, keeping the PMOS of the following gate active, increasing its duty cycle, worsening NBTI aging of that gate by Equation~\ref{eq:gatedelay}.
For efficiency, ATPG tools produce patterns which test many gates at once. Choosing patterns for effective TWAs simply means to select patterns which cover the most faults along the target path. It follows from Equation~\ref{eq:accelerationfactor} that increasing the PMOS duty cycle for as many gates as possible on this path accelerates its NBTI-induced aging.
\sethlcolor{yellow}


\textit{2) Formal Methods.}
As just stated,
to accelerate NBTI-induced aging along a path means to increase the PMOS duty cycle for as many gates as possible on that path.
Toward that end, attackers can formulate cover properties that require the input pins of gates in the path to remain stable for many cycles.
Note that further constraints could be added to specify which gates should be held ON/1 vs which OFF/0, to enforce as many PMOS transistors being ON as possible.
However, exploring all possible combinations -- and even more so formulating and feeding the related properties to the tools -- for ON vs OFF assignments for all the gates along the path may not be scalable.
Therefore, it is more practical to formulate and use only the initial constraints for stable states, then generate many compliant input patterns with low computational cost, and finally select patterns that also do hold as
many PMOS transistors ON as possible.


\subsubsection{Devising the Attack Code}

In \ding{204}, C code or hand-written assembly code is devised following the selected input patterns and their bit vectors.
Each of these vectors can be translated into a micro-operation packet or instruction, which shall contain the matching opcode, operands, and any relevant control
signals.
In other words, these instructions place the functional unit into the same operational state as described by the bit vectors.
Note that the actual instructions are also simulated against both the gate-level netlist, for verification of aging, and the architecture, for correctness; related details are provided in Section~\ref{sec:CS:setup}.

In the code, the resulting set of instructions is repeatedly called to accelerate aging.
First, however, the generated operands are placed into dedicated registers,
to avoid unnecessary memory requests between instructions.
Likewise, to avoid introducing additional latency due to register dependencies, these instructions all write to some register not occupied by the operands.

\subsubsection{Attack Deployment}

In \ding{205}, the attack code is deployed to a victim system.
As the unit under attack stays perpetually busy with the attack code, the gates of the target path experience high duty cycles, so NBTI-induced aging steadily degrades performance of that path
(Equation~\ref{eq:gatedelay}).
For any shared or multi-threaded system, once the operating system would place some victim process on the same core, faulty computations would occur once the malicious wearout kicks in.
Note that the actual time until wearout can be determined by aging-aware STA; see Section~\ref{sec:CS} for details.



\section{Case Study}
\label{sec:CS}

For a practically relevant case study, we chose the FPU of the Berkeley Out-of-Order Machine (BOOM)~\cite{Celio:EECS-2015-167}.
Here, we refine the general methodology for TWA with all technical details as needed. In particular, we utilize industry standard design tools along with
aging-characterized technology libraries and standard guardbands.


\subsection{Implementation and Setup of Attack}
\label{sec:CS:setup}

\subsubsection{Synthesis, Timing Analysis, and Path Selection}

The BOOM core uses the Berkeley Hardfloat IP~\cite{hauserBerkeleyHardFloat} for its FPU.
Without loss of generality, the FPU was isolated and synthesized with Synopsys Design Compiler using the NanGate 14nm FinFET PDK~\cite{silvacoSilvacoSi2Release2019}. The latter was accurately modeled by
Klemme et al.~\cite{klemmeEfficientLearningStrategies2022}. 
STA was performed using Synopsys PrimeTime against the same PDK/libraries
for timing closure at 1.25 GHz.
Note that, in a real-world attack scenario, an attacker would skip synthesis and perform STA with their reverse engineered netlist.

\begin{figure}
    \centering
    \includegraphics[width=.6\linewidth]{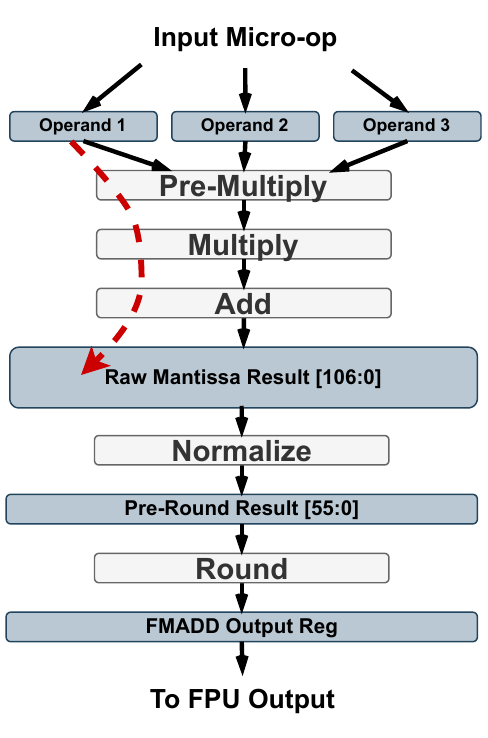}
    \caption{The BOOM FPU's fused multiply-add datapath. The exemplary path under attack is outlined by the dashed arrow.}
    \label{fig:fmadd_pipe}
\end{figure}

Using the timing information obtained from STA, a near-critical path was chosen among the double-precision fused multiply-add datapaths that support the RISC-V FMADD instruction family.
Outlined in Figure~\ref{fig:fmadd_pipe}, this path comprises 25 gates spanning from bit 35 of the first operand to bit 101 of the unrounded, unshifted mantissa.
This is an effective path to attack, as the final exponent of a floating-point result is determined by the number of leading 0s in the mantissa before it is shifted to the left for the normalized representation.
That is, these leading 0s are subtracted from the exponent produced by the calculation to generate the final exponent.
Thus, a timing violation here would most likely induce a larger exponent value and an incorrectly shifted mantissa.  Of course should an attacker desire a more subtle or more extreme attack, different paths that are also near-critical could be chosen.



\subsubsection{Generation of Aging-Inducing Attack Patterns}

Recall that we employ and explore two complementary methods here, using ATPG tools vs using formal methods.

\textit{1) ATPG.}
NBTI-induced aging accelerates when PMOS transistors are held ON continuously.
To find input patterns that induce this behavior, we used Synopsys TestMAX ATPG~\cite{synopsysincTestMAXATPGAdvanced}
and configured it to test for stuck-at-1 faults in the design,
\sethlcolor{green}
as discussed in Section \ref{sec:patterngeneration}.
\sethlcolor{yellow}
For an effective attack, the patterns covering the most stuck-at faults for the gates along the targeted
path were selected.

We also feed the patterns to the same DUT for functional simulation using Synopsys VCS, for the following reason.  First, note that ATPG supports only structural testing which, in contrast to functional testing, is
agnostic to the functionality of the DUT, in particular to functional details such as opcodes. Thus, including ATPG-generated patterns directly as is for instructions would mean to risk adding idle cycles for the FPU,
generating unimplemented instruction exceptions, etc., which would all disrupt the execution of the attack.
Thus, we filter out all
the generated patterns whose op-codes did not correspond to valid RISC-V FMADD instructions.
For simplicity, we kept special floating-point values such as NaNs (not-a-number) and infinity
in the patterns as-is. This is practical without risking attack interruptions,
as RISC-V does not trap upon raising a floating-point exception flag (but rather relies on software to check these flags~\cite{risc-vinternationalRISCVInstructionSet}).

From over 900 patterns generated, we select two exemplary patterns, with IDs 82 and 176, respectively.
\sethlcolor{green}
These patterns covered the largest number of stuck-at faults along the target path, 14 and 26, respectively.
\sethlcolor{yellow}
Pattern 82 had 3 invalid instructions removed, and Pattern 176 had 6 removed.
Both patterns had 7 valid FPU instructions for practical use in the attack.

\textit{2) Formal Methods.}
Here, the Cadence JasperGold proof engine "L" was used to generate the input vector set.
Engine "L" was chosen for its heuristic-driven traversal of the design's state space~\cite{cadencedesignsystemsincJasperEngineSelection2024}, which effectively mutates the inputs as it searches, avoiding trivial solutions whenever possible.

Through SystemVerilog assumptions, we constrained the input space to valid RISC-V FMADD operations, with accordingly set micro-op control bits.
Since RISC-V propagates NaN and infinity values provided as operands directly to the FPU outputs (without passing them through the computational paths),
constraints were added to prevent NaNs and infinity from being generated as operands. Note that this is in contrast to the ATPG approach.
To ensure the FPU's data paths would actually be utilized, further constraints were added to prevent the generation of simple moves (e.g. FMV, FLD, FSD) or other instructions such as FMIN, FMAX, or FCLASS.
Finally, a cover property was crafted to describe the desired state for the FPU.
This property,
effectively enforces that the five specified nets must remain unchanged/stable for 10 cycles.


Once the engine reached this covered state, the resulting input trace was simulated and confirmed that these nets were indeed stable and, thus, the trace was directly usable for an attack.
Without loss of generality, 10 cycles was chosen as cover length to obtain input traces of similar length to those obtained from ATPG.

Of the 20 traces generated by JasperGold accounting for all the constraints, four were eventually selected as candidates for the attack:
traces 0, 1, 3, and 4 were selected as, out of 25 gates in the path in total, these traces held the largest number of PMOS transistors ON, namely at least 12.


\subsubsection{Devising the Attack Code}
\label{sec:atkprog}

Next, we transform the selected patterns into a binary executable for the target system.
As indicated, constructing an actual program to perform TWA was a matter of splitting each input vector in a pattern into its opcode bitfield and the operand(s).
From there, RISC-V assembly instructions are written out, removing any instructions whose opcode is invalid or irrelevant to the target unit.
For our implementation, a representative subset of which is shown below in Listing~\ref{lst:asm}, the instructions were placed in a long running loop and the operands were loaded into the RISC-V floating-point registers beforehand.

\footnotesize
\begin{lstlisting}[label={lst:asm},caption=ATPG Pattern 176 as RISC-V Assembly Code,basicstyle=\small]
attack():
        lui     t4,%hi(.operand0)
        fld     ft4,%lo(.operand0)(t4)
        lui     t3,%hi(.operand1)
        fld     ft3,%lo(.operand1)(t3)
        lui     t1,%hi(.operand2)
        fld     ft2,%lo(.operand2)(t1)
        lui     a7,%hi(.operand3)
        ...
.Loop:
        fmul.d ft5, ft4, ft3
        fmul.d ft5, ft2, ft1
        fmsub.d ft5, ft0, fa0, fa1
        fclass.d a5, fa2
        fsub.d ft5, fa3, fa4
        fcvt.w.s a5, fa5, rtz
        j       .Loop
main:
        addi    sp,sp,-16
        sd      ra,8(sp)
        call    attack()
.operand0:
        .word   2081935025
        .word   -67742208
.operand1:
        ...
\end{lstlisting}
\normalsize

\begin{figure*}[ht!]
    \centering
    \includegraphics[width=.8\linewidth]{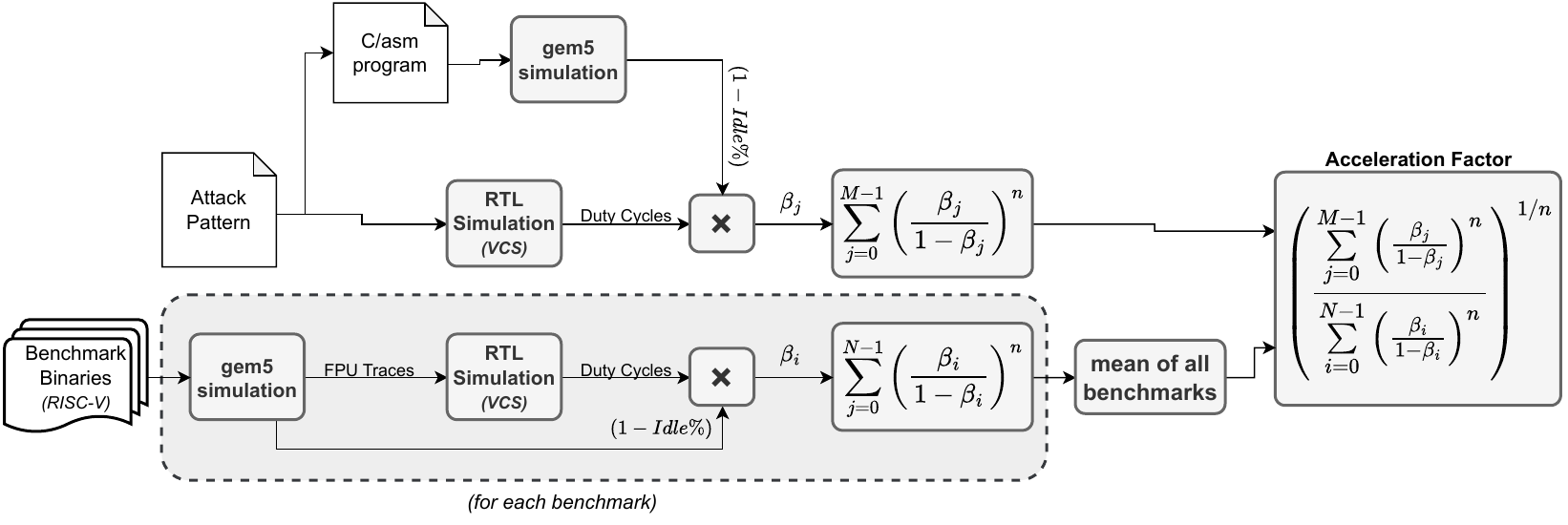}
    \caption{Evaluation flow for TWA, focused on the functional correctness of the attack code and the acceleration of aging triggered by the attack code until wearout.
    }
    \label{fig:evaluationflow}
\end{figure*}

\begin{figure*}[ht!]
\begin{minipage}[b]{0.32\textwidth}
    \centering
    \includegraphics[width=\textwidth]{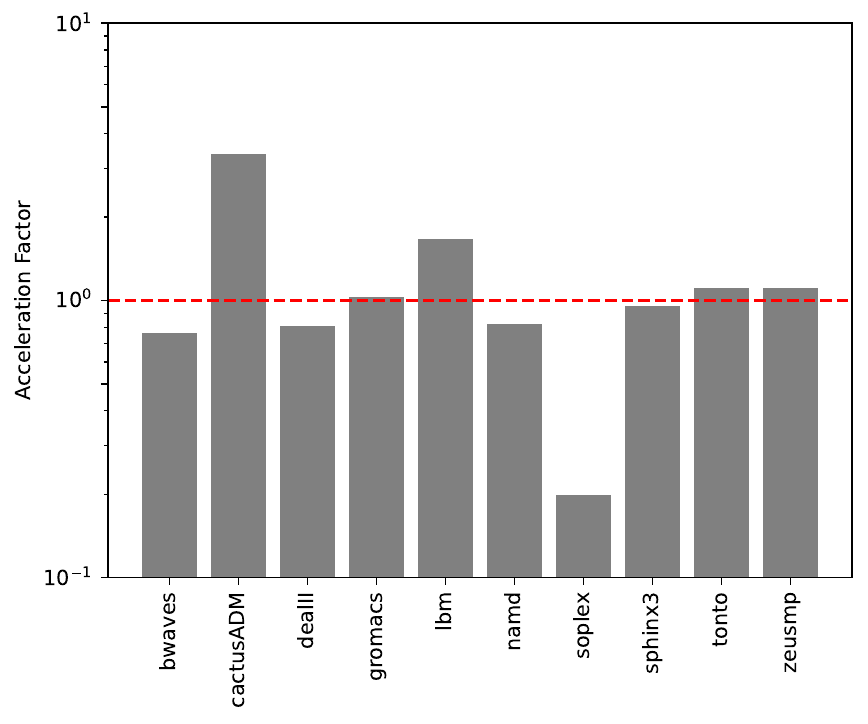}
    \caption{Relative AF for the SPECfp2006 benchmarks, with respect to the mean $T_{lifetime}$.
    }
    \label{fig:AF_spec}
\end{minipage}
\hfill
\begin{minipage}[b]{0.32\textwidth}
    \centering
    \includegraphics[width=\textwidth]{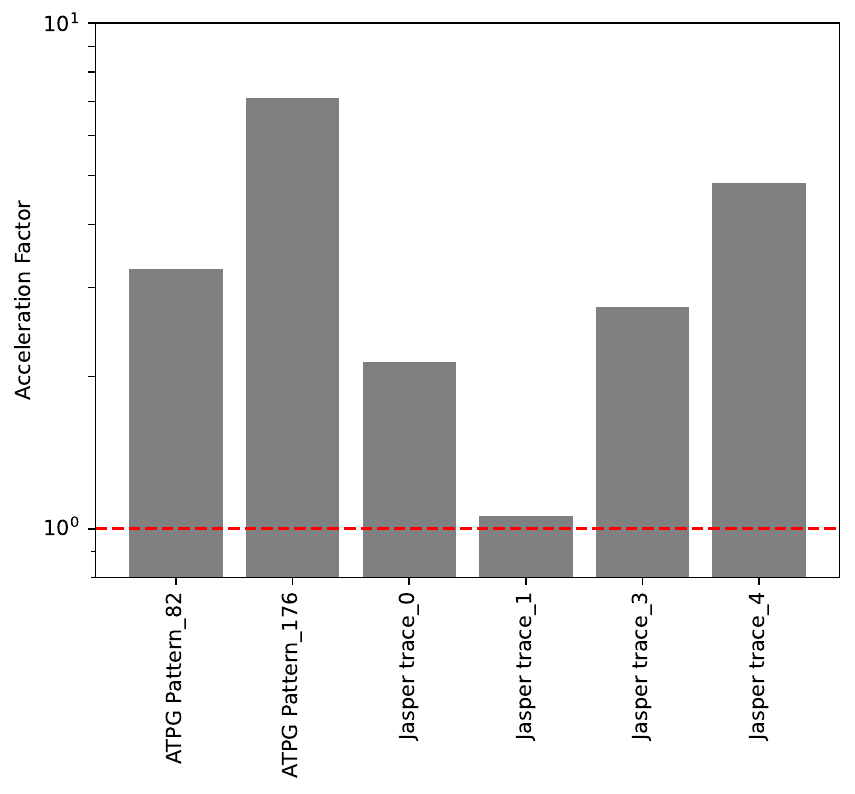}
    \caption{AFs for the tested TWA attack patterns.}
    \label{fig:AF_attack}
\end{minipage}
\hfill
\begin{minipage}[b]{0.32\textwidth}
    \centering
    \includegraphics[width=\textwidth]{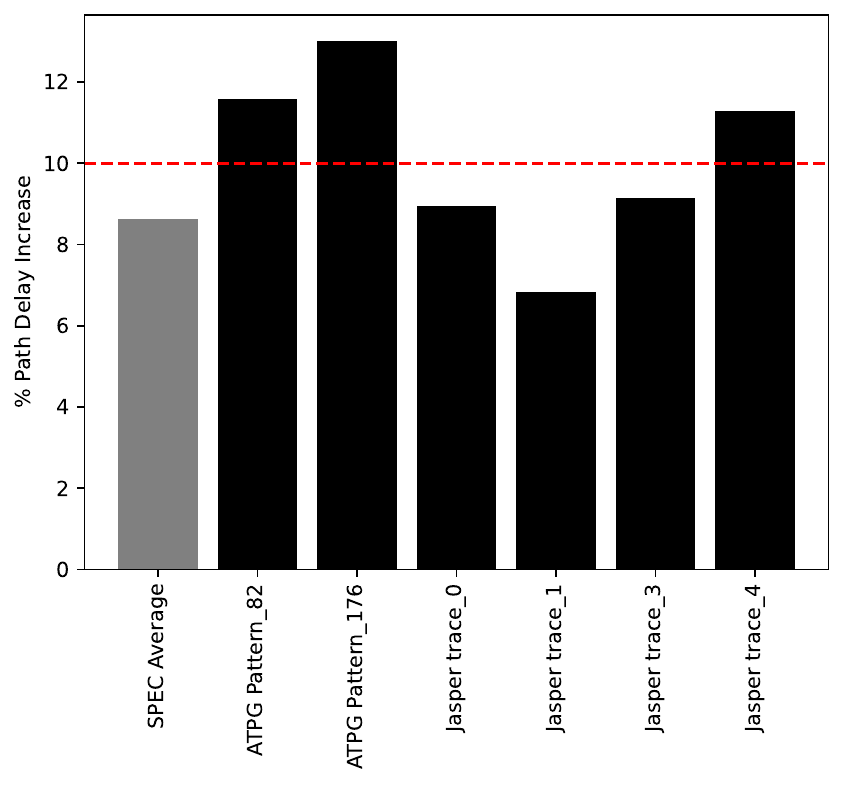}
    \caption{Path delay degradation demonstrated using an ML-characterized standard cell library}
    \label{fig:delay}
\end{minipage}
\end{figure*}


	The code in Listing~\ref{lst:asm} was also simulated in the gem5 architectural simulator~\cite{binkertGem5Simulator2011,lowe-powerGem5SimulatorVersion2020}.
	This was done to determine the percentage of cycles the FPU was not issued an instruction, which was 14\%.
	We reduce the duty cycles accordingly (by the corresponding percentage points), to account for idle times, before applying them to Equation~\ref{eq:accelerationfactor}.
	This follows the premise that any cycles in which we are not utilizing the FPU are not contributing to aging, thereby establish an more appropriate lower bound for the accelerated aging.
	More details for the evaluation of TWA are described next.

\subsection{Evaluation Setup}

The evaluation setup is illustrated in Figure~\ref{fig:evaluationflow} and described next.

\subsubsection{Baseline Wear}

To assess the wearout induced by the proposed TWA, we first have to determine some nominal/baseline wear.
%
Toward that end, without loss of generality, we determine FPU activities for the SPEC2006FP benchmark suite as follows.

SimPoints~\cite{hamerlySimPoint30Faster2005} of each benchmark were taken. To capture the timing, micro-operation codes, and the operand data being dispatched to the FPU, the RISC-V64G compiled binaries
were simulated in gem5.
These traces were simulated on the RTL model using Synopsys VCS.
Then, the baseline aging of the system was obtained by capturing the duty cycles for each gate along the target path, across each
benchmark (see dashed grey box in Figure~\ref{fig:evaluationflow}). All duty cycles were adjusted for idle time in the same manner as applied to the attack's duty cycles (Section~\ref{sec:atkprog}).
The denominator sum in Equation~\ref{eq:accelerationfactor} was calculated for each benchmark and the mean was taken as the reference aging for comparison with the attack.

	
\subsubsection{Acceleration Factor for Malicious Aging Until Wearout}
\label{sec:calculateAF}


We assume that the CPU under attack is designed and manufactured to operate at the designated frequency with a timing guardband large enough that, during nominal wear, correct operation is ensured for a lifetime of 5 to 10 years~\cite{moghaddasiAgingAwareInstructionLevelStatistical2020,kimAgingGracefullyApproximation2019a}.
Accordingly, we say that, for any Acceleration Factor (AF) induced by the attack that is exceeding this nominal wear, the time until wearout (i.e., the reduction in operational lifetime until failure) is
accelerated~\cite{kimUseItLose2013}.
For example, should our attack enable an AF of 5x over baseline wear, and assuming the circuit and its guardband are designed for a 5-year nominal lifetime, then our attack would induce errors in
the targeted path after one year of accelerated wear.

As described above, all the attack patterns' duty cycles were adjusted for idle time
before applying Equation~\ref{eq:accelerationfactor} to determine the AF of the NBTI-induced, maliciously triggered aging. The top of Figure~\ref{fig:evaluationflow} depicts this process.
Note that other factors such as temperature remain unchanged across the baseline and the attack setting.

\subsubsection{Verification of Timing Violations}

To confirm each pattern's efficacy for TWAs, we utilize a set of aging-characterized libraries corresponding to the baseline 14nm FinFET PDK.
Timing is characterized for two years of aging, under varying duty cycles from 0 to 100\%, and for 85C.
For further details on the aging characterization process, refer to Klemme et~al.~\cite{klemmeMachineLearningCircuit2021}.

For any attack pattern, the netlist was independently customized by replacing all the gates that experienced attack-specific aging with the corresponding cells from the set of duty-cycle-dependent libraries.
These revised netlists then underwent STA and timing was compared with the baseline netlist.

\subsection{Results}
\label{sec:results}

Figure~\ref{fig:AF_spec} illustrates the relative aging rates for each SPEC2006 benchmark tested, compared to the arithmetic mean of the set of $T_{lifetime}$ values. There are two outlier benchmarks responsible for the standard deviation of 0.86. Soplex has a rather low AF of 0.199, owing to the long recorded periods in which the FPU remains idle, measured using the gem5 simulator. cactusADM was the other notable benchmark, whose consistent utilization of the fused multiply-add logic yielded an AF of 3.39.

The AF for each tested attack pattern relative to the mean aging of the SPEC benchmarks is shown in Figure~\ref{fig:AF_attack}. From the methods discussed in Section~\ref{sec:TWAgeneralflow}, two ATPG patterns and four JasperGold patterns were selected as candidates for an attack. ATPG pattern 176 produced the attack with the strongest NBTI aging acceleration. This pattern had a large number of gates with high duty cycles, and following Equation~\ref{eq:accelerationfactor}, this produced an aging AF of 7.1X, effectively cutting the target path's mean time to failure by that factor.

The aging acceleration by the JasperGold patterns was more mixed than those of the ATPG patterns. Trace 1 aged the path only a little more than the SPEC benchmarks. This pattern had the most gates whose input duty cycles were 0, contributing nothing toward NBTI aging.

\subsection{Analysis}
\label{sec:analysis}

\subsubsection{Demonstrated Impact of TWA Attack}

\sethlcolor{green}
The best-case AF suggests the attack could succeed in 8 months, for a device designed to operate for 5 years, as a lower-bound for aging acceleration, as noted in Section~\ref{sec:atkprog}.
\sethlcolor{yellow}
 ATPG Pattern 176's performance degradation was confirmed with STA on the aged standard cell libraries~\cite{klemmeMachineLearningCircuit2021}, which modeled aging after two years. Using STA with these libraries.  The results of the analyses for all of the tested attack patterns are shown in Figure~\ref{fig:delay}.  The figure shows the strongest attack's path delay degradation to be 13.0\%, significantly more than the presumed 10\% guardband. 

\begin{figure}
    \centering
    \includegraphics[width=.7\linewidth]{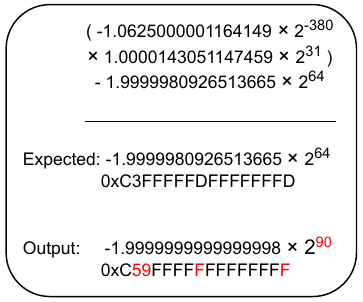}
    \caption{Erroneous calculation caused by TWA attack
    }
    \label{fig:casestudy}
\end{figure}

With the timing information from STA on Pattern 176, the FPU was simulated with generated FMADD tests from the Berkeley TestFloat program~\cite{hauserBerkeleyHardFloat}. An example calculation is shown in Figure~\ref{fig:casestudy}. As seen in the figure, both the mantissa and the exponent of the result are incorrect.
In floating-point arithmetic, after the calculation operation has been performed, the mantissa is not yet shifted to a normal 1.xxxxxx representation. In the pipeline shown in Figure~\ref{fig:fmadd_pipe}, the FPU normalization stage counts the leading zeros in
the mantissa to determine how far left to shift the mantissa bits, and this count is subtracted from the intermediate exponent value.

So, in this attack, an errant 1 at bit 103 of the intermediate mantissa reduces this count by 26, altering the final exponent to 0b10001011001 instead of 0b10000111111, hence the difference in exponent. The two altered mantissa bits are more simply explained. The most near-critical paths in this design were in the cones of input to the upper intermediate mantissa bits. Many of the gates targeted in this attack were shared among those nearby paths, and so their timing degraded also, to varying extents. As discussed in Section~\ref{sec:logicalimpact}, timing violations occur when old data latches because the gate paths have become too slow to propagate the new data in time. For reference, the previous output of this operation was 0x419FFFFFFDFFEFFF.

\begin{figure}
    \centering
    \includegraphics[width=.7\linewidth]{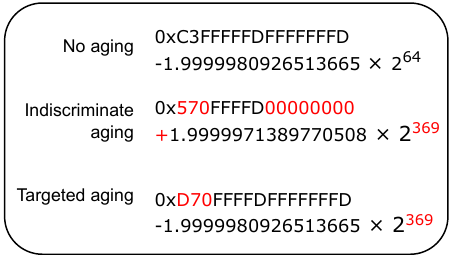}
    \caption{Example FPU calculation highlighting the precision of the attack}
    \label{fig:aging_compare}
\end{figure}

The precision of the attack is demonstrated in Figure~\ref{fig:aging_compare}. In this example obtained from simulation, the previous calculation produced an output of 0 to better illuminate bits whose paths became slow. Here, in a device whose aging advanced more evenly (middle), half of the mantissa failed to propagate and the sign bit also missed timing, in addition to the error in the exponent seen with the targeted aging.
Here in the example, we see that compared to more aggressive, indiscriminate aging, the targeted aging effect is more discreet.

\subsubsection{Thermal Estimation of the Attack}
\begin{figure}
    \centering
    \includegraphics[width=.7\linewidth]{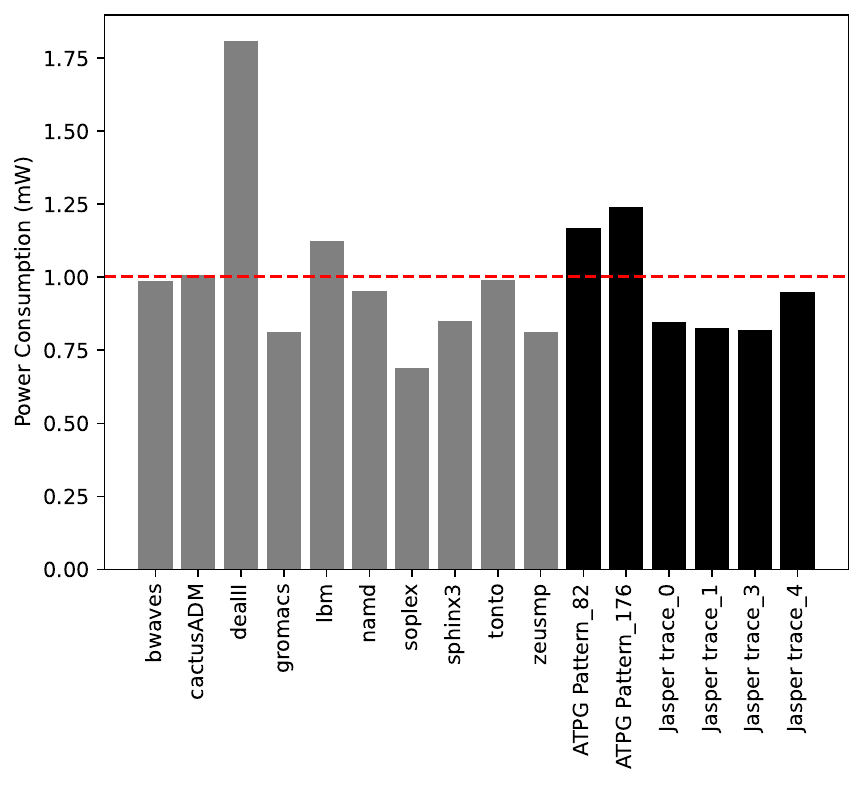}
    \caption{Power consumption estimates for SPEC2006 and attack traces}
    \label{fig:power}
\end{figure}

A condition for the AF calculation to be an effective predictor of aging is that while under baseline and test (attack) conditions, the thermal characteristics should be similar or perhaps greater, to worsen the aging (Equation~\ref{eq:gatedelay}).

To investigate, we used Synopsys PrimePower to characterize the estimated power consumption on the FPU using the activity captured during RTL simulation, since increased switching activity dissipates more heat, accelerating aging via Equation~\ref{eq:gatedelay}. The results are seen in Figure~\ref{fig:power}.
The horizontal dashed line is the average power consumption among the SPEC benchmarks. To allow the study to focus on the impacts of increased PMOS duty cycles on NBTI aging of the target path, we establish that the attack traces achieve at least a normal level of overall activity in the FPU. The ATPG-derived attack patterns perform well in this regard, while the patterns from JasperGold generate less activity, but still align with the middle of the SPEC benchmarks (side-stepping testing solutions which observe unit switching activity such as FitBit~\cite{swaminathanFitBitEnsuringRobust2024}). This is a side effect of using formal methods on their own, which tend to take the path of least resistance to cover properties. This could be improved by adjusting the random mutation parameters in the formal engine. That these traces were optimized to hold selected gates at a high input duty cycle, a state not conducive to high power consumption, suggests that attacks which optimize both for PMOS duty cycles and for increased switching of surrounding gates would prove more effective.

\subsubsection{Comparing the two attack generation methods}
Both the ATPG and formal engine methods achieved some measure of success in crafting attack patterns that accelerated aging in a targeted path. Automatic Test Pattern Generation, when considering the AFs, power estimates, and delay increases due to Patterns 82 and 176, 
     yielded the most effective attack vectors. However, ATPG tools can be quite slow; our generation of 900 patterns took about two days on an AMD EPYC 7443 using 16 cores. 
     Additionally, the patterns needed to be cross-referenced with the chosen path for overlap with each pattern's list of testable faults. Furthermore, the patterns generated did not necessarily constitute a valid
     executable attack. In the above experiment, any vector in the pattern whose opcode was invalid or not a floating-point operation was discarded, which still yielded seven valid instructions from both patterns. Pattern 82 only had three instructions cut, but Pattern 176 had six cut. However, methods to constrain the pattern space during ATPG would provide better control over the instruction sequence.

Using a solver such as one in Cadence JasperGold hypothetically yields a much more specific attack due to the nature of SystemVerilog assertions. The runtime is also much lower, with tens of patterns being found in
minutes. Using assumption statements, the operations and the nature of the operands can be easily controlled. However, with these added constraints, many patterns may need to be generated before an effective attack is found. The choice of solver engine has a large impact on the randomness of the inputs. For a desirable attack, the inputs should differ enough to induce more switching activity across the unit while holding the gates under attack stable. JasperGold's "L" engine incorporates some fuzzing logic which tries to mutate the inputs, which makes it an ideal choice compared to engines which strictly aim for Boolean satisfiability, which for our purposes, would produce patterns such as: the inputs do not change for 10 cycles, since that's the trivial solution to keeping gate input stable.

\section{Mitigation Possibilities and Future Work}
\label{sec:disc}

We present this work to raise awareness in the microarchitecture community, enabling timely development of defenses. In this section, we outline several initial ideas toward effective mitigation.
While NBTI stress degrades the performance of CMOS circuits, the effects are not necessarily permanent. When the PMOS transistor is no longer negatively-biased, the device is in a \textit{relaxation} state, where some of
the trapped charge carriers regain mobility and $V_{th}$ can reduce somewhat~\cite{alamComprehensiveModelPMOS2007,grasserParadigmShiftUnderstanding2011,bhardwajPredictiveModelingNBTI2006}. It follows, then, that reducing the duty cycles of PMOS transistors can reduce NBTI aging.
A common approach for doing this without exacerbating PBTI effects is using exercise vectors, sets of input bit vectors stored in a ROM that are deployed to mitigate NBTI stress and elongate the lifespan of the device~\cite{kimUseItLose2013,vitkovskiyClothoProactiveWearout2015,pendyalaGateLevelNBTI2020,reddyMinimalExerciseVector2017}.

Such input vectors could possibly be deployed to mitigate TWA attacks. Without loss of generality, functional units in a CPU could be equipped with logic to deploy exercise vectors during idle periods between heavy use. The question then becomes how to initiate this sequence often enough to mitigate intentional wearout without inflating power consumption or accelerating other modes of aging such as hot-carrier injection.
Circuit-level monitors~\cite{kimAgingGracefullyApproximation2019a} could be deployed here to intelligently initiate an exercise sequence~\cite{qiNBTIResilientCircuits2008,synopsysincSLMPathMargin}, with varying area overhead. Also pertinent is that the TWA methodology could be reapplied to different designs, perhaps different functional units on the same CPU. The effort, area, power, and performance overhead of monitoring and mitigating units across a CPU core may quickly become costly.

Efforts to detect TWA attacks could be made at a higher abstraction level, such as adding counters to a CPU core to monitor for behavior patterns that seem malicious. Take for example the attack program in Listing~\ref{lst:asm}, where specially crafted operands are preloaded from memory into registers before a loop of independent instructions executes. The pattern of repeated writes to registers whose values are never used could
certainly be tracked with counters set to trigger mitigation patterns when a threshold is reached. However, this is not the only way to deploy such an attack, and one skilled enough in assembly language optimization
could devise means to work around such pattern detection, exemplified by the ongoing efforts to mitigate the RowHammer DRAM attack~\cite{mutluFundamentallyUnderstandingSolving2023}. Finding an architectural solution to mitigate TWA attacks and extend the operating lifetime of CPU cores is an ongoing effort by the authors.

\sethlcolor{green}
Further ongoing work by the authors aims to improve the efficacy and speed of the TWA attack. Two objectives include refining the formal-methods-based generation of attack patterns and exploring co-optimization strategies that combine NBTI acceleration with increased switching activity to raise local temperature, thus amplifying the overall aging effect.
\sethlcolor{yellow}

\section{Conclusion}
\label{sec:concl}
Negative-bias temperature instability is an ongoing reliability and security concern in complex nanoscale CMOS circuits. In this work, we introduced the \textbf{Targeted Wearout Attack}, a method by which NBTI aging in a chosen logic path in a CPU core could be accelerated by a specially crafted program running without elevated privileges. We suggested and compared methods to devise the attack program and followed up with a case study on a modern RISC-V CPU. We used both the Acceleration Factor and an aging-characterized standard cell library to quantify the aging acceleration of the demonstrated attack, which reduced the functional lifetime of the targeted path by a factor of over 7.1. Finally, we suggested possible approaches and trade-offs for mitigation and lifetime extension.



\bibliographystyle{IEEEtranS}
\bibliography{references}

\end{document}